\documentclass[english]{article}
\usepackage[T1]{fontenc}
\usepackage[latin9]{inputenc}
\usepackage{amsmath}
\usepackage{amssymb}
\usepackage{graphicx}

\makeatletter
\newcommand{\lyxaddress}[1]{
\par {\raggedright #1
\vspace{1.4em}
\noindent\par}
}
\newenvironment{lyxlist}[1]
{\begin{list}{}
{\settowidth{\labelwidth}{#1}
 \setlength{\leftmargin}{\labelwidth}
 \addtolength{\leftmargin}{\labelsep}
 }}
{\end{list}}

\usepackage{babel}

\newcommand{\ket}[1]{|#1 \rangle}

\newcommand{\sandwich}[3]{\left \langle #1 \mid #2 \mid #3 \right\rangle}

\usepackage{babel}

\usepackage{babel}

\usepackage{babel}

\usepackage{babel}

\makeatother

\usepackage{babel}
\begin{document}

\title{Using the J1-J2 Quantum Spin Chain as an Adiabatic Quantum Data Bus}

\author{Nick Chancellor$^{1}$ and Stephan Haas$^{1}$}

\maketitle

\lyxaddress{1 Department of Physics and Astronomy and Center for Quantum Information
Science \& Technology, University of Southern California, Los Angeles,
California 90089-0484, USA}
\begin{abstract}
This paper investigates numerically a phenomenon which can be used
to transport a single q-bit down a J1-J2 Heisenberg spin chain using
a quantum adiabatic process. The motivation for investigating such
processes comes from the idea that this method of transport could
potentially be used as a means of sending data to various parts of
a quantum computer made of artificial spins, and that this method
could take advantage of the easily prepared ground state at the so
called Majumdar-Ghosh point. We examine several annealing protocols
for this process and find similar results for all of them. The annealing
process works well up to a critical frustration threshold. There is
also a brief section examining what other models this protocol could
be used for, examining its use in the XXZ and XYZ models. 
\end{abstract}

\section*{Introduction}

The ability to send data from one part of a computer to another accurately
and quickly is an essential feature in virtually any design. The use
of artificial spin clusters in quantum computing has been of growing
interest. There is an implementation which has been demonstrated using
superconducting flux q-bits\cite{Johnson2011,Harris2010-1,Perdomo2008,vanderPloeg2006,Harris2010-2}.
This paper demonstrates an effective and scalable way of sending arbitrary
q-bit states along a spin chain with Heisenberg type coupling using
quantum annealing. Assuming one could implement a Hamiltonian which
follows this model, for example using the methods proposed in \cite{Chen2010}
using coupled cavities, this system design could be used for a data
bus which transports quantum states to different sections of a quantum
computer system. For instance, the protocols discussed in this paper
could potentially be used to move states from memory to a system of
quantum gates in an implementation of the circuit model.

There has already been significant work done on the subject of quantum
data buses using spin chains, \cite{Banchi2010,Banchi2011,Apollaro2012,Banchi2011-2}.
However these works differ significantly from the method proposed
in this paper in that the encoded q-bit is not transmitted through
a degenerate ground state manifold, but through excitations of the
Hamiltonian.

This paper investigates a method of using q-bits as an intermediate
bus for the transfer of quantum information. This method can be compared
to another method which is that of pulses \cite{Fitzsimmons2006},
where a Hamiltonian is applied to a system for a period of time to
perform a given operation. In the case of information transfer this
operation is usually a swap. Unlike the method of using pulses, this
method of using q-bits does not require precise timing to insure that
the correct operation is performed. The method of using a spin chain
Hamiltonian as a data bus also means that one does not need to either
be able to address any pair of q-bits in the system or perform multiple
operations to transfer an arbitrary q-bit. The pulse method does have
the advantage that every intermediate spin can be used as quantum
memory. However this is at the cost of the increased complexity of
using dynamic quantum evolution in excited states, and the requirement
of precise timing.

The adiabatic quantum bus method also has the advantage that, as in
any adiabatic quantum process, only the lowest energy parts of Hamiltonian
need to be faithfully realised by the implementation method. For example,
a Hamiltonian which actually has an infinite number of excited states
on each ``spin'', but where only the low energy states which act
like a spin $\frac{1}{2}$ Heisenberg system, contribute to the ground
state would be perfectly acceptable to use as an adiabatic quantum
bus without modification. But the higher energy states may cause issues
using a method such as pulses. This general feature of adiabatic quantum
processes such as the one illustrated in this paper makes them more
versatile than their non-adiabatic counterparts.

The effect we will examine exploits the SU(2) symmetry of the Heisenberg
Hamiltonian and uses the ground state degeneracy created by this symmetry
in a chain with an odd number of spins. It has already been demonstrated
\cite{Chancellor2011} that disturbances can be sent an unlimited
distance along such chains because of their degenerate ground state.
This paper goes a step further and actually demonstrates how a specific
state can be transported across the chain using a quantum annealing
protocol. Further investigation will also be provided into application
of this method to systems such as the XYZ spin chain which only have
a $\mathbb{Z}_{2}$symmetry.

\subsection*{Setup\label{sub:Setup}}

The model we consider is the J1-J2 Heisenberg spin chain with open
boundaries,

\begin{equation}
H=\sum_{n=1}^{N-1}J_{1}^{n}\vec{\sigma}_{n}\cdot\vec{\sigma}_{n+1}+\sum_{n=1}^{N-2}J_{2}^{n}\vec{\sigma}_{n}\cdot\vec{\sigma}_{n+2}.\label{eq:J1-J2}
\end{equation}

This model has SU(2) symmetry, which is expressed by the Hamiltonian
being block diagonal, such that there are N+1 blocks each with $\binom{N}{k}$
states. Each block represents all of the states with a given number,
k, of up spins. If the number of spins in the model is odd, then the
additional symmetry under a flip of the spins in the z direction,
i.e. $\sigma^{z}\rightarrow-\sigma^{z}$ implies that all states of
the Hamiltonian have a twofold energy degeneracy. In the anti-ferromagnetic
case, ( $J_{1},\: J_{2}>0$ ) the ground state manifold consists of
one state from the $\textrm{k=floor(}\frac{N}{2}\textrm{)}$ and one
from the $\textrm{k=ceil(}\frac{N}{2}\textrm{)}$ sector. A simple
example of this would be taking a system with 5 spins, the ground
state would be twofold degenerate and would span the k=2 and k=3 sectors.
One can now consider an initial Hamiltonian of the form of Eq. \ref{eq:J1-J2}
where the couplings are the ones given by

\begin{eqnarray}
J_{1}^{n} & = & \begin{cases}
J_{1}^{n,init} & n<N-1\\
0 & n=N-1
\end{cases},\label{eq:J1 init}\\
J_{2}^{n} & = & \begin{cases}
J_{2}^{n,init} & n<N-2\\
0 & n=N-2
\end{cases}.\label{eq:J2 init}
\end{eqnarray}

The general condition on $J_{1}^{n,init}$ and $J_{2}^{n,init}$ is
that the coupling is predominantly anti-ferromagnetic everywhere and
that each spin is coupled to the others by at least one non zero J.
For simplicity this paper considers only $J_{1}^{n,init}=1$ and $J_{2}^{n,init}=J_{2}^{init}$.
This ground state manifold consists of the tensor product of the (unique)
ground state of the chain of length N-1 with the Nth spin in an arbitrary
state, a state in this manifold is of the from given by

\begin{equation}
\ket{\Psi^{init}}=\ket{\Psi_{0}^{N-1}}\times\ket{\psi^{init}},\label{eq:gs. init}
\end{equation}

where $\ket{\Psi_{0}^{N-1}}$ is the ground state of the spin chain
of length N-1 and $\ket{\psi^{init}}$is an arbitrary single spin
state. One can now consider the same Hamiltonian, but with $n\rightarrow(N-n)+1$
. This Hamiltonian also has the form of Eq. \ref{eq:J1-J2}, but with
couplings

\begin{eqnarray}
J_{1}^{n} & = & \begin{cases}
J_{1}^{n,final} & n>1\\
0 & n=1
\end{cases},\label{eq:J1 fin}\\
J_{2}^{n} & = & \begin{cases}
J_{2}^{n,final} & n>2\\
0 & n=2
\end{cases}.\label{eq:J2 fin}
\end{eqnarray}

The general condition on $J_{1}^{n,final}$ and $J_{2}^{n,final}$
is that the coupling is predominantly anti-ferromagnetic everywhere
and that each spin is coupled to the others by at least one non-zero
J. For simplicity this paper considers only $J_{1}^{n,final}=1$ and
$J_{2}^{n,final}=J_{2}^{final}$. A state in the ground state manifold
is now given by

\begin{equation}
\ket{\Psi^{final}}=\ket{\psi^{final}}\times\ket{\Psi_{0}^{N-1}},\label{eq:gs. fin}
\end{equation}

where $\ket{\psi^{final}}$ is an arbitrary single spin state. One
can now consider a quantum annealing process with described by

\begin{equation}
\textrm{\textrm{H}(t;}\tau\textrm{)}=\textrm{\textrm{\textrm{A}(t;}\ensuremath{\tau}\textrm{)} H}^{\textrm{init}}+\textrm{\textrm{\textrm{B}(t;}\ensuremath{\tau}\textrm{)} H}^{\textrm{final}},\label{eq:H anneal}
\end{equation}

where $\textrm{H}^{\textrm{init}}$ is \ref{eq:J1-J2} with the conditions
given in \ref{eq:J1 init} and \ref{eq:J2 init} and $\textrm{H}^{\textrm{final}}$
is \ref{eq:J1-J2} with the conditions given in \ref{eq:J1 fin} and
\ref{eq:J2 fin}. Also A and B follow the conditions 
\begin{eqnarray}
\textrm{\textrm{A}(}t\leq0\textrm{;}\tau\textrm{)} & = & 1,\label{eq:anneal conds. begin}\\
\textrm{\textrm{B}(}t\leq0\textrm{;}\tau\textrm{)} & = & 0,\\
\textrm{\textrm{A}(}t\geq\tau\textrm{;}\tau\textrm{)} & = & 0,\\
\textrm{\textrm{B}(}t\geq\tau\textrm{;}\tau\textrm{)} & = & 1.\label{eq:anneal conds. end}
\end{eqnarray}

\begin{figure}
\includegraphics[scale=0.5]{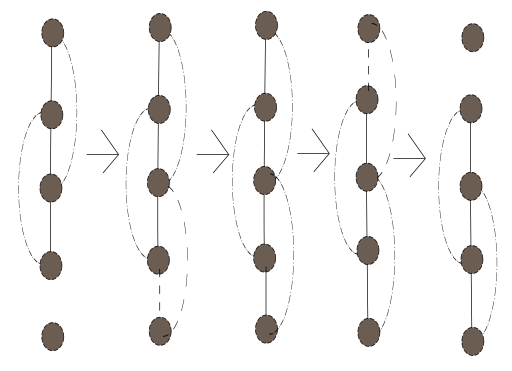}

\caption{Cartoon representation of a process where a spin is joined to the
chain, then the spin on the opposite end is removed. Note that this
is only one specific example of many possible processes for transporting
a q-bit.\label{fig:transport cartoon}}
\end{figure}

For all values of A and B the SU(2) symmetry is preserved. Therefore
the Hamiltonian remains block diagonal at all times. The symmetry
of the Hamiltonian under $\sigma^{z}\rightarrow-\sigma^{z}$ is also
preserved at all times. This implies that the ground-state degeneracy
(as well as the twofold degeneracy of all states) is preserved. The
block diagonal structure implies that there will be no exchange of
amplitude between spin sectors during the annealing process, while
the degeneracy implies that no relative phase can be acquired between
the states in the $\textrm{k=floor(}\frac{N}{2}\textrm{)}$ and the
$\textrm{k=ceil(}\frac{N}{2}\textrm{)}$ sector. From the combination
of these two conditions one can see that as long as one anneals slowly
enough with $\textrm{\textrm{H}(t;}\tau\textrm{)}$ %
\footnote{Technically one must give the additional condition that there is no
true crossing within the spin sectors on the annealing path.%
} one can start with a state of the form given in Eq. \ref{eq:gs. init}
and reach a final state in the form Eq. \ref{eq:gs. fin} where $\ket{\psi^{fin}}=\exp(\imath\varphi)\ket{\psi^{init}}$,
and $\varphi$ is an irrelevant phase. One specific example of such
an annealing protocol to transport a spin is given in Fig. \ref{fig:transport cartoon}.

\section*{Advantages}

The use of the J1-J2 Heisenberg chain for transport by quantum annealing
has several advantages. First the model with uniform coupling is gapped
for $\frac{J_{2}}{J_{1}}\gtrsim0.25$ \cite{Chitra1995}. This suggests
that within the adiabatic evolution process, at least locally, the
system should behave as a gapped system in this regime, as long as
global effects such as odd length frustration do not cause problems.

It is important to note that even the largest system size considered
here is far from the thermodynamic limit. One should note, however,
that given the connectivity schemes of adiabatic quantum chips already
in existence \cite{Harris2010-2}, one may only need to transport
a q-bit state a few spins to get it to any part of the system.

Further evidence of favorable scaling comes from \cite{Chancellor2011}
which demonstrates that disturbances can travel an unlimited distance
in the presence of a degenerate ground state, even in a gapped system.
Furthermore, \cite{Chancellor2011} suggests that these disturbances
can carry entanglement, polarization, and quantum information. The
transport by annealing given here is a specific example of how this
effect can be taken advantage of.

Another advantage of the use of the J1-J2 Heisenberg Hamiltonian is
the existence of the so called Majumdar-Ghosh point \cite{Majumdar1970}
($\frac{J_{2}}{J_{1}}=0.5$). At this point the ground state (with
an even number of spins) has the simple form of a matrix product of
singlets. Due to this fact the system should be relatively easy to
prepare. The system is also gapped at the Majumdar-Ghosh point, making
the Majumdar-Ghosh Heisenberg Hamiltonian, an ideal system for transport
by quantum annealing and the ideal candidate for building an adiabatic
quantum data bus.

Although this paper focuses on the J1-J2 Heisenberg model, it should
be noted that this same annealing scheme should work with any pattern
of coupling in the intermediate spins (i.e. J1-J2-J3)%
\footnote{At least this should work for small systems. In the continuum limit
many of these systems may become gapless, so that quantum annealing
cannot be effectively performed. Also one may be able to construct
certain pathological cases with paths which pass through true crossings.%
}. One would also expect this scheme to work in models where the SU(2)
symmetry is broken but there is a remaining $\mathbb{Z}_{2}$symmetry
such as the XYZ or XY model, again with arbitrary patterns of coupling.
Note however that this mehhod will not work in the Ising model, because
although there is a $\mathbb{Z}_{2}$ symmetry, the Hamiltonian lacks
terms to exchange q-bits between sites because it is diagonal in the
computational basis.

\section{Proof of Principle}

None of the arguments so far have given much illumination to the difficulty
or ease of annealing within the sector. While we have discussed that
transport of a q-bit state is possible in principle by annealing,
we have not yet shown that the annealing process is fast enough to
be practical. For this we turn to numerics. For the purposes of this
paper we will consider the annealing time, $\tau$ , required to reach
a given fixed fidelity, $F(\tau)$, with the true final ground state,

\begin{equation}
F(\tau)=|\sandwich{\Psi^{fin}}{\intop_{0}^{\tau}dt\, H(t,\tau)}{\Psi^{init}}|.\label{eq:anneal time}
\end{equation}

The J1-J2 Heisenberg model is not an analytically solved model, at
least for finite values of $J_{2}$, so numerical methods must be
used in this calculation. One can first consider one part of the annealing
process, in which a single spin is joined to a even length J1-J2 spin
chain, using both $J_{1}$ and $J_{2}$ couplings which are linearly
increased to equal values of those used in the rest of the chain %
\footnote{Note that this Hamiltonian (and all other annealing Hamiltonians in
this paper) can be rewritten in the form of \ref{eq:H anneal}. However
it is much more compact not to write the unchanging parts of the Hamiltonian
twice.%
},

\begin{equation}
H(t,\tau)=\sum_{n=1}^{N-2}J_{1}\vec{\sigma}_{n}\cdot\vec{\sigma}_{n+1}+\sum_{n=1}^{N-3}J_{2}\vec{\sigma}_{n}\cdot\vec{\sigma}_{n+2}+\lambda(t,\tau)(J_{1}\vec{\sigma}_{N-1}\cdot\vec{\sigma}_{N}+J_{2}\vec{\sigma}_{N-2}\cdot\vec{\sigma}_{N}),\label{eq:join}
\end{equation}
 
\[
\lambda(t,\tau)=\begin{cases}
0 & t\leq0\\
\frac{t}{\tau} & 0<t<\tau\\
1 & t\geq\tau
\end{cases}.
\]

\begin{figure}
\includegraphics[scale=0.4]{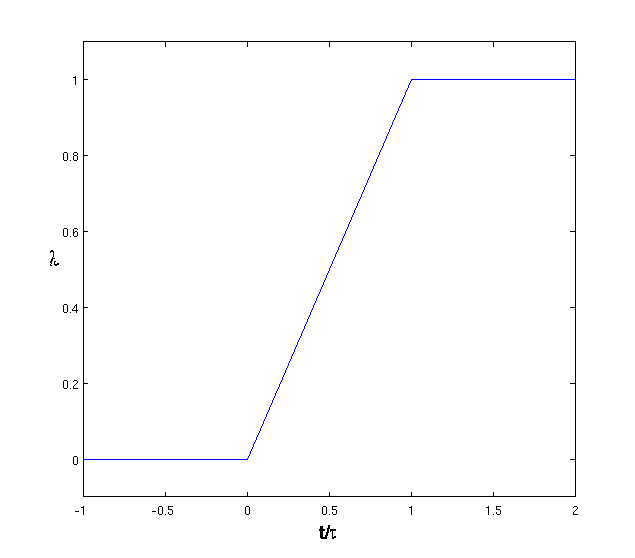}

\caption{Coupling constant $\lambda(t,\tau)$ from Eq.\ref{eq:join} and Eq.
\ref{eq:teleport H} versus $\frac{t}{\tau}$ .}
\end{figure}

\begin{figure}
\includegraphics[scale=0.4]{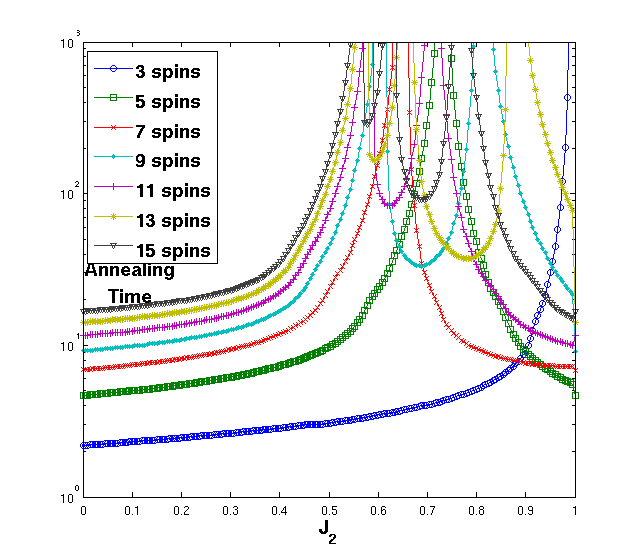}

\caption{Annealing time required to reach a 90\% fidelity with the true ground
state within one of the two largest spin sectors of the Hamiltonian
vs. $J_{2}$, with $J_{1}$ set to unity. One can see that for larger
values of $J_{2}$ the annealing time behaves unpredictably. The annealing
time also scales poorly with system size close to the Majumdar-Ghosh
point.\label{fig:T_join}}
\end{figure}

As shown in Fig. \ref{fig:T_join}, the annealing time required becomes
large and highly sensitive to small variations for larger values of
$J_{2}$. Also the behavior seems to get worse in this regime as system
size is increased, and is poor at the Majumdar-Ghosh point %
\footnote{At least for fixed coupling, the case of dynamically changing coupling
will be considered later.%
}.

\begin{figure}
\begin{lyxlist}{00.00.0000}
\item [{\includegraphics[scale=0.4]{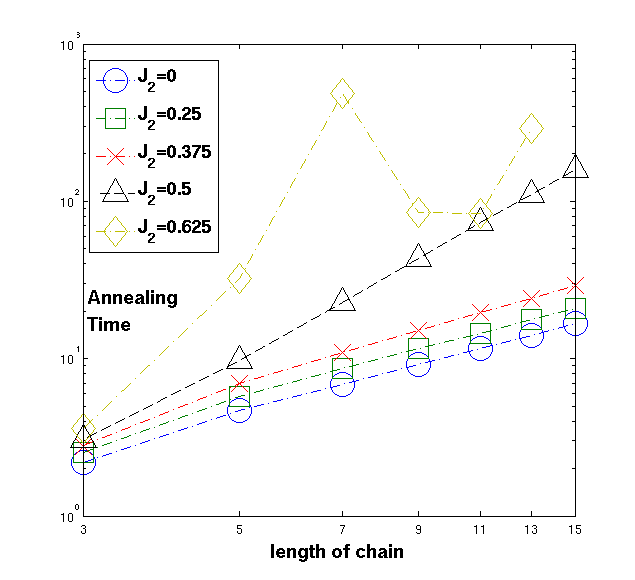}}]~
\end{lyxlist}
\caption{\label{fig:time_scaling} Scaling of annealing time to achieve 90\%
final ground state fidelity (in units of inverse Hamiltonian energy)
versus length of chain on a log-log plot. }
\end{figure}

As a further demonstration of the scaling with annealing time versus
$J_{2}$, one can plot the annealing time versus system size, as we
have done in Fig. \ref{fig:time_scaling}. This figure shows polynomial
or even sub polynomial scaling for small values of $J_{2}$, but than
shows strongly non-monotonic behavior for stronger coupling. It is
important to note however that even the longest chain length considered
here is probably far from the infinite system limit, and this data
may not be trustworthy for making predictions for scaling as the chain
length approaches the infinite system limit.

\begin{figure}
\includegraphics[scale=0.4]{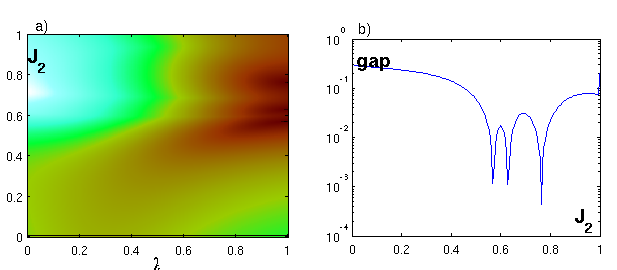}

\caption{\label{fig:join_gap} Plots of gap for joining a single spin to an
even length J1-J2 Heisenberg spin chain. For density plots lighter
colors indicate larger gap. a) gap versus $\lambda$ in Eq. \ref{eq:join}
and $J_{2}$ for 15 total spins d) Gap versus $J_{2}$ with $\lambda=1$}
\end{figure}

By examining the gap one can hope to gain insight into the underlying
cause of the behaviour of annealing time curves. As Figs. \ref{fig:join_gap}(a)
and (b) show, the behavior of the annealing time curves is reflected
by the presence of what appear to be true crossings %
\footnote{Strictly speaking nothing in this paper has demonstrated them to be
true crossings, they could just be close avoided crossings, it does
not matter for the purposes of this paper.%
} for the odd length spin chain with uniform coupling. Fig. \ref{fig:join_gap}(b)
shows the gap for an odd length spin chain and seems to confirm the
presence of points with very small gap with uniform coupling for $J_{2}$
above 0.5. Figs. \ref{fig:T_join} and \ref{fig:join_gap} together
show that, at least at the length scales considered here, there are
good annealing paths for joining a single spin to an even length chain.
However, the simplest method of taking advantage of the simple ground-state
wavefunction at the Majumdar-Ghosh point is not optimal. Fortunately
there are many other possible options to take advantage of the easily
prepared ground state and hopefully avoid the regions of small gap
found here.

\section{Dynamically Tuning J2}

One method to avoid regions of small gap while still taking advantage
of the Majumda-Ghosh point would be to start at the Majumdar-Ghosh
point and then dynamically reduce the value of $J_{2}$ during the
annealing process, a simple way of doing this would be to use the
Hamiltonian in Eq. \ref{eq:dynamic J2}.

\begin{equation}
H(t,\tau)=\sum_{n=1}^{N-2}J_{1}\vec{\sigma}_{n}\cdot\vec{\sigma}_{n+1}+\sum_{n=1}^{N-3}J_{2}(t,\tau)\vec{\sigma}_{n}\cdot\vec{\sigma}_{n+2}+\lambda(t,\tau)(J_{1}\vec{\sigma}_{N-1}\cdot\vec{\sigma}_{N}+J_{2}(t,\tau)\vec{\sigma}_{N-2}\cdot\vec{\sigma}_{N}),\label{eq:dynamic J2}
\end{equation}

\[
\lambda(t,\tau)=\begin{cases}
0 & t\leq0\\
\frac{t}{\tau} & 0<t<\tau\\
1 & t\geq\tau
\end{cases},
\]

\[
J_{2}(t,\tau)=\begin{cases}
0.5 & t\leq0\\
0.5+\frac{t}{\tau}(J_{2f}-0.5) & 0<t<\tau\\
J_{2f} & t\geq\tau
\end{cases}.
\]

\begin{figure}
\includegraphics[scale=0.4]{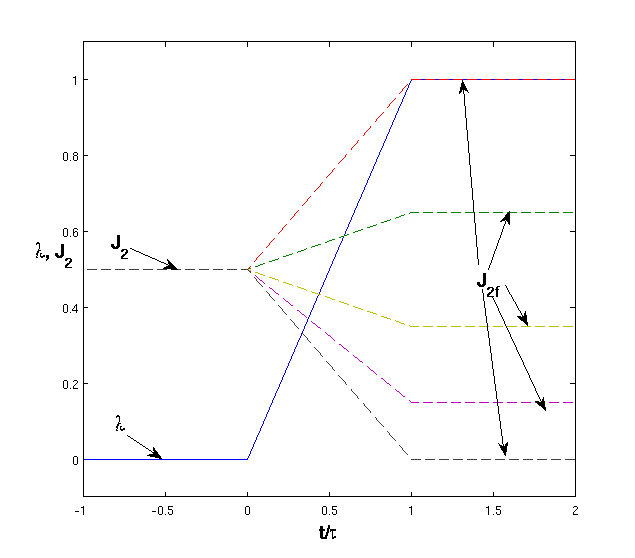}\caption{In this annealing protocol not only is a spin coupled to the chain,
but $J_{2}$is also changed dynamically.}
\end{figure}

\begin{figure}
\includegraphics[scale=0.4]{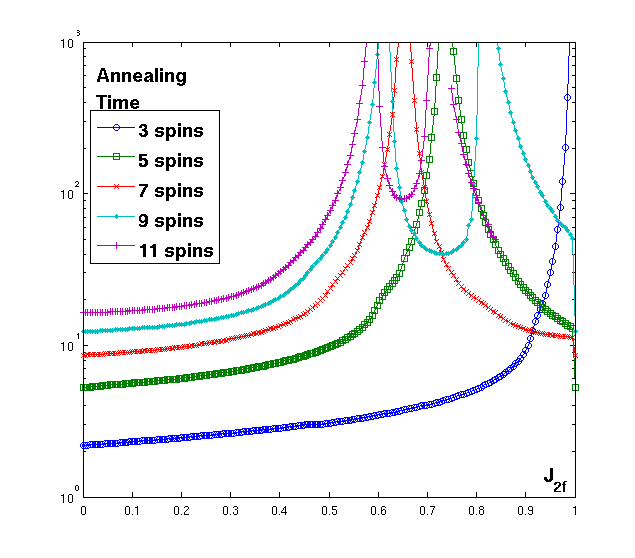}

\caption{\label{fig:MDG_time} Annealing time required to reach a 90\% fidelity
with the true ground state within one of the two largest spin sectors
of the Hamiltonian with dynamical coupling starting at $J_{2}$=0.5
and linearly changing to $J_{2f}$ while also joining a spin to the
chain, with $J_{1}$ set to unity throughout the process. Notice that
this figure is qualitatively and quantitatively very similar to Fig.
\ref{fig:T_join}. }
\end{figure}

Fig. \ref{fig:MDG_time} shows that taking advantage of the easily
prepared ground state at the Majumdar-Ghosh point does in fact work,
and the curves in this figure are strikingly similar to those in Fig.
\ref{fig:T_join}. This similarity is to be expected because Fig.
\ref{fig:join_gap} demonstrates that the gap is the smallest where
the spin is completely joined. Hence this part of the process should
dominate the annealing time.

It is reasonable to argue that because the regions of phase space
which are visited are the same in the uncoupling process as coupling,
the behavior of the system during the uncoupling process is determined
by the gaps shown in Fig. \ref{fig:join_gap}, and therefore the annealing
times for the uncoupling process should be at least qualitatively
similar to those given in Fig. \ref{fig:T_join}. One advantage to
the uncoupling process is that unlike the coupling process, the need
is not as strong to end in an easily prepared state. The only reason
one may have to want to end in the Majumdar-Ghosh point is as an error
check. The spins in the chain can be measured after the end of the
process to ensure that no error has occurred %
\footnote{For example if two spins which should be in a singlet together ended
up being measured to be facing in the same direction than the annealing
process would have failed.%
}.

\begin{figure}
\includegraphics[scale=0.4]{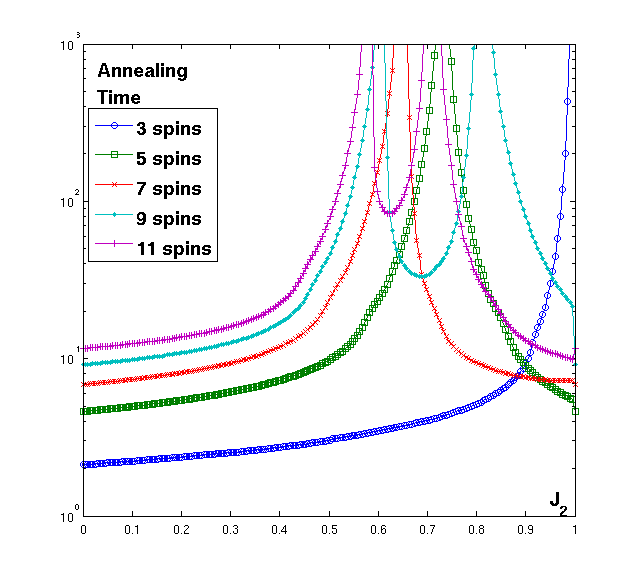}

\caption{\label{fig:uncoupling_time}Annealing time required to reach a 90\%
Fidelity with the true ground state for uncoupling process within
one of the two largest spin sectors of the Hamiltonian vs. $J_{2}$
with $J_{1}$ set to unity. One can see that this figure is very similar
to Fig. \ref{fig:T_join} as one would expect because it is simply
the time reversed version of that process.}
\end{figure}

Fig. \ref{fig:uncoupling_time} shows the time required to uncouple
a spin from the chain, not surprisingly this figure looks very similar
to Fig. \ref{fig:T_join} which is the coupling process. Note that
in this system the Hamiltonian is simply Eq. \ref{eq:join} with $\frac{t}{\tau}\rightarrow(1-\frac{t}{\tau})$
.

\begin{figure}
\includegraphics[scale=0.4]{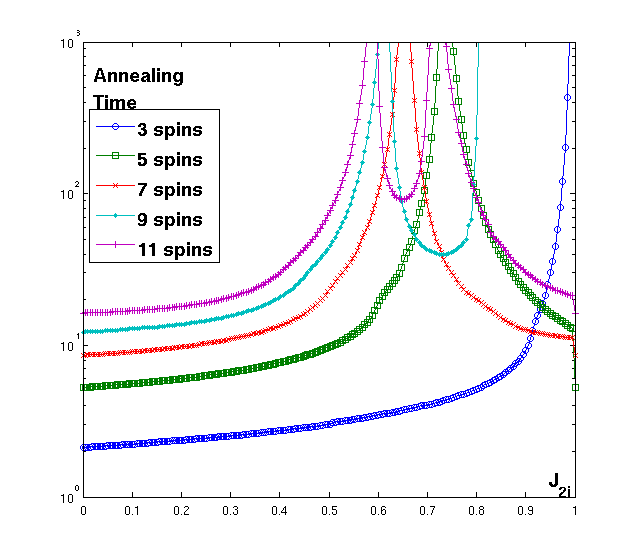}

\caption{\label{fig:uncoupling_MDG}Annealing time required to reach a 90\%
fidelity with the true ground state for uncoupling process within
one of the two largest spin sectors of the Hamiltonian vs. initial
$J_{2i}$ with a final $J_{2}$ at the Majumdar-Ghosh point with $J_{1}$
set to unity. This figure is very similar to Fig. \ref{fig:MDG_time}
as one would expect, because it is simply the time reversed version
of that process. }
\end{figure}

As expected, except for one curve where a numerical error made some
points unable to plot one can see from Fig. \ref{fig:uncoupling_MDG}
that the uncoupling process also requires roughly the same time as
the coupling process for dynamically tuned $J_{2}$. Note that the
Hamiltonian for this process is simply Eq. \ref{eq:dynamic J2} with
$\frac{t}{\tau}\rightarrow(1-\frac{t}{\tau})$ and $J_{2f}\rightarrow J_{2i}$.

\section{Simultaneous Uncoupling and Coupling}

Because many of the issues encountered with the coupling protocol
seem to relate to odd-spin frustration, it may be reasonable to consider
simultaneously coupling one q-bit to the chain while uncoupling the
other. The Hamiltonian in this case is given in Eq. \ref{eq:teleport H}.

\begin{equation}
H(t,\tau)=\sum_{n=1}^{N-2}J_{1}\vec{\sigma}_{n}\cdot\vec{\sigma}_{n+1}+\label{eq:teleport H}
\end{equation}

\[
\sum_{n=1}^{N-3}J_{2}(t,\tau)\vec{\sigma}_{n}\cdot\vec{\sigma}_{n+2}+\lambda(t,\tau)((J_{1}\vec{\sigma}_{N-1}\cdot\vec{\sigma}_{N}+J_{2}\vec{\sigma}_{N-2}\cdot\vec{\sigma}_{N})-(J_{1}\vec{\sigma}_{1}\cdot\vec{\sigma}_{2}+J_{2}\vec{\sigma}_{1}\cdot\vec{\sigma}_{3})),
\]

\[
\lambda(t,\tau)=\begin{cases}
0 & t\leq0\\
\frac{t}{\tau} & 0<t<\tau\\
1 & t\geq\tau
\end{cases}.
\]

\begin{figure}
\includegraphics[scale=0.4]{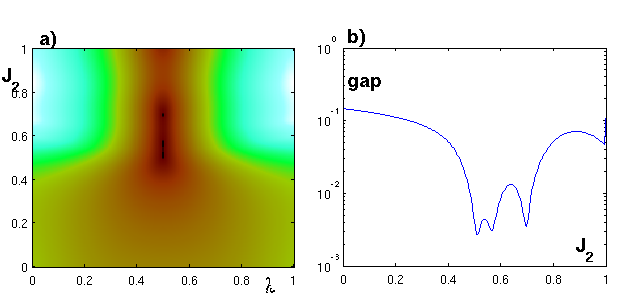}

\caption{\label{fig:teleport_gaps}Plots of gap for simultaneously joining
a single spin to an even length J1-J2 Heisenberg spin chain and unjoining
a spin from the other end. For density plots lighter colors indicate
larger gap. a) gap versus $\lambda$ from Eq. \ref{eq:teleport H}
and J2 for 17 total spins b) Gap versus $J_{2}$ with $\lambda=0.5$.}
\end{figure}

Fig. \ref{fig:teleport_gaps} shows the gaps for various system sizes
for the process where the couplings are turned on and off simultaneously.
This process does not seem to avoid the area of low gap for $J_{2}\gtrsim0.5$
seen in Fig. \ref{fig:join_gap}. However by comparing Fig. \ref{fig:teleport_gaps}
d) and Fig. \ref{fig:join_gap} d) one can see that it appears that
the process of simultaneous uncoupling and coupling is characterized
by avoided crossings rather than true crossings %
\footnote{This statement is based on the fact that the gap does not have a cusp
when plotted on a log scale. Strictly speaking this just shows that
there is not a true crossing at the line where the two couplings are
equal.%
}.

\begin{figure}
\includegraphics[scale=0.4]{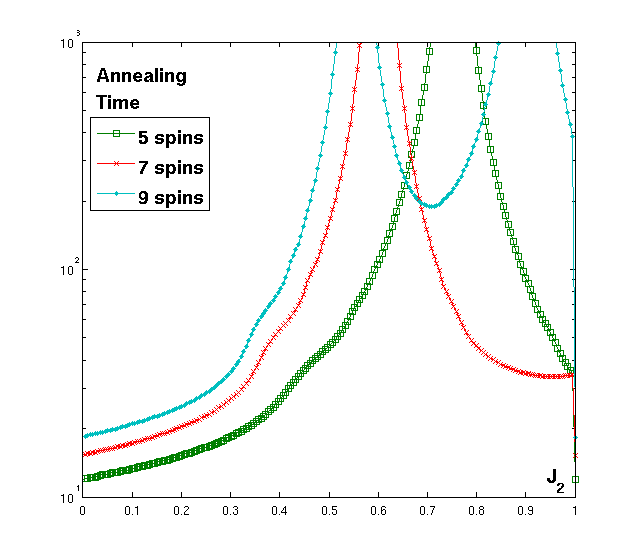}

\caption{\label{fig:teleport time}Annealing time required to reach a 90\%
Fidelity with the true ground state for combined coupling and uncoupling
process within one of the two largest spin sectors of the Hamiltonian
vs. $J_{2}$ with $J_{1}$ set to unity. }
\end{figure}

Fig. \ref{fig:teleport time} shows the time required for annealing
processes with for the combined coupling and uncoupling process, the
results are consistent with what one would expect from looking at
Fig. \ref{fig:teleport_gaps}, and confirm that the annealing time
also tends to be very long and vary a lot for larger values of $J_{2}$.

\section{Requirements for use as an Adiabatic Quantum Bus}

It is now useful to consider a broader class of models that may be
used as adiabatic quantum buses, as in general the full SU(2) symmetry
of the Heisenberg Hamiltonian is not required.

The requirements for a spin chain (or network) Hamiltonain to be usable
as an adiabatic quantum bus are as follow: 
\begin{enumerate}
\item The ground state must be at least 2 fold degenerate, and the ground
state manifold must be able to encode a q-bit. In this paper this
is achieved by having at least a $\mathbb{Z}_{2}$symmetry, and an
odd number of spins, but there may be other ways. 
\item The Hamiltonian (or at least the low energy states) must be predominantly
anti-ferromagnetic in nature. This guarantees that the encoded q-bit
will be excluded from the larger spin chain (or network) when a single
q-bit is removed. 
\item The Hamiltonian must contain terms which perform exchanges between
sites. This excludes models such as the Ising model which, although
it has the required symmetry, cannot be used a quantum bus because
its Hamiltonian is diagonal in the computational basis 
\item One must be able to slowly couple in a spin with an arbitrary state
on one end of the chain (network) and also to slowly remove coupling
on the other end. More control may improve performance, but is not
necessary. 
\item Annealing paths in parameter space must not contain true crossings.
This is a general requirement for adiabatic quantum computing. 
\end{enumerate}

\section{XXZ and XYZ model}

As previously mentioned, the full SU(2) symmetry of the Heisenberg
Hamiltonian is not required. The Hamiltonian must only have a $\mathbb{Z}_{2}$symmetry
to encode and transport one q-bit of information. In this section
we will briefly examine two other possibilities: the XXZ model, where
the SU(2) symmetry is broken, but the block diagonal structure imparted
by this symmetry remains, and the XYZ model where only the block diagonal
structure of a $\mathbb{Z}_{2}$ symmetry is present.

\begin{figure}
\includegraphics[scale=0.4]{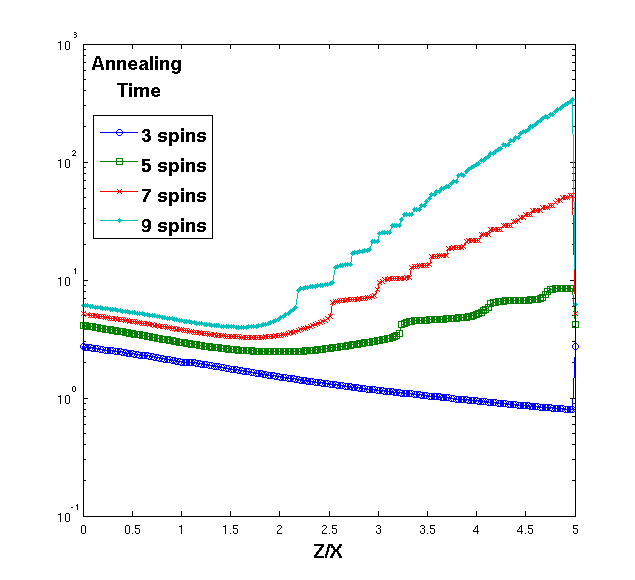}

\caption{\label{fig:XXZ} Annealing time to reach 90\% fidelity on using the
adiabatic quantum bus protocol on an XXZ spin chain versus the ratio
of X and Z coupling strengths note that Z/X=0 is an XX model while
Z/X=1 is a J1 Heisenberg spin chain. This data was obtained with joining
and disconnecting of spins occurring simultaneously.}
\end{figure}

As one can see from Fig. \ref{fig:XXZ}, the XXZ model can be used
as an adiabatic quantum data bus. There is a regime where this system
outperforms the XXX Heisenberg model for Z/X between 1 and roughly
2. This is to be expected because adding additional coupling in the
z direction may serve to open the gap between the the ground-state
manifold and the next excited state. The increasing time as the z
coupling is increased further can be explained because the system
would behave like an Ising model in the limit of $\frac{Z}{X}\gg1$
.

One can further examine the behavior of an XYZ model as an adiabatic
quantum spin bus. For this purpose we consider the quantum bus protocol
performed on the following normalized XYZ Hamiltonian

\begin{equation}
H_{XYZ}(\Delta;N)=C_{\Delta}\sum_{i=1}^{N-1}\sigma_{i}^{x}\sigma_{i+1}^{x}+(1+\Delta)\sigma_{i}^{y}\sigma_{i+1}^{y}+(1+2\Delta)\sigma_{i}^{z}\sigma_{i+1}^{z},\label{eq:XYZ_h}
\end{equation}

where the normalization is

\[
C_{\Delta}=\frac{\sqrt{3}}{\sqrt{1+(1+\Delta)^{2}+(1+2\Delta)^{2}}}.
\]

One can now examine the performance of this Hamiltonian for different
values of $\Delta$, noting that $H_{XYZ}(0;N)$ is simply the J1
Heisenberg spin chain of length N.

\begin{figure}
\includegraphics[scale=0.4]{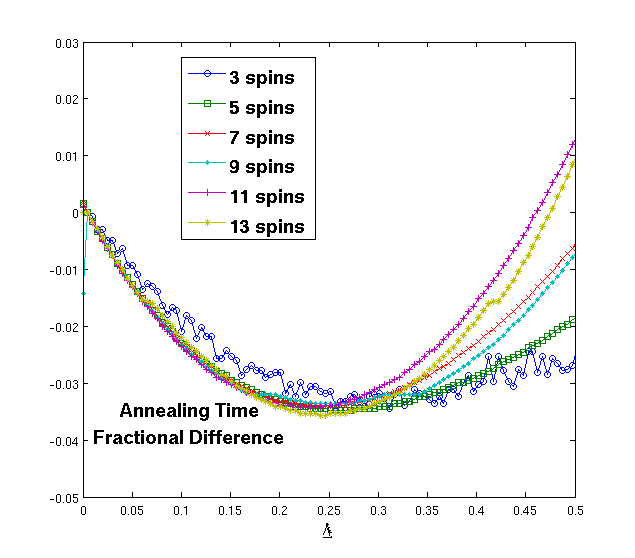}

\caption{\label{fig:XYZ_plot} Plot of fractional difference from annealing
time for an chain with small $\Delta$ (Heisenberg chain). This data
is for the adiabatic quantum bus protocol performed on a chain of
the form eq. \ref{eq:XYZ_h} with spins being attached and removed
simultaneously.}
\end{figure}

As Fig. \ref{fig:XYZ_plot} shows, a slight advantage can be gained
by using an XYZ model rather than a simple Heisenberg chain. Fig.
\ref{fig:XYZ_plot} also seems to suggest that the benefit gained
is relatively independent of chain length.

\section{Other Protocols}

So far we have only investigated a small subset of the possible annealing
protocols which meet the criteria given in the introduction. For example
the XY spin chain should also have and easily prepared ground state
and may be easier to experimentally realize \cite{Johnson2011}. One
could also try to examine the case of dynamically tuning the y and
or z direction coupling and starting out at the Majumdar-Ghosh point
but using modified coupling in the y and z directions with an XYZ
model to avoid low gap regions.

One could also try to change the coupling scheme to avoid the low
gap region, by either randomly or systematically modifying the coupling
between intermediate spins, if this is done dynamically, one can still
take advantage of the Majumdar-Ghosh point. This technique could also
be used in conjunction with any of the ideas in the previous paragraph.

This paper is intended only to provide proof of principle for this
method and is by no means an exhaustive search of all possible protocols.

\section{Conclusions}

We have demonstrated how a J1-J2 Heisenberg spin chain can be used
to transport a q-bit state adiabatically. We have also shown that
many extensions of this Hamiltonian; such as different coupling schemes
or the XY or XYZ model which have only a $\mathbb{Z}_{2}$ symmetry,
will also be able to be used to transport a q-bit %
\footnote{Assuming there is not a true crossing along the annealing path, the
coupling must also be (at least predominately) anti-ferromagnetic
so that the excess spin does not become trapped in the larger spin
chain.%
}. We have found that for values of high frustration, transport by
quantum annealing does not work very well. We have also demonstrated
that this does not prevent us from exploiting the easily prepared
ground state at the Majumdar-Ghosh point. We have given some examples
of possible annealing protocols in this paper, but have really only
investigated a very small section of a vast space of possible protocols
for transportation of quantum states by annealing.

\section*{Acknowledgements}

The numerical computations were carried out on the University of Southern
California high performance supercomputer cluster. This research is
partially supported by the ARO MURI grant W911NF-11-1-0268.

\end{document}